\documentclass[12pt,preprint]{aastex}

\usepackage[usenames,dvipsnames]{color}
\usepackage{ulem}

\begin{document}


\title{Integrated Optical Polarization of Nearby Galaxies}


\author{Amy Jones}
\author{Lifan Wang} 
\author{Kevin Krisciunas} 
\author{Emily Freeland}
\affil{George P. \& Cynthia W. Mitchell Institute for Fundamental Physics and
Astronomy,\\ Department of Physics and Astronomy, Texas A\&M University, College
Station, TX 77843, USA}
\email{ymamay@physics.tamu.edu}


\begin{abstract} 

We performed an integrated optical polarization survey of $70$ nearby galaxies to
study the relationship between linear polarization and galaxy properties.  To date
this is the largest survey of its kind.  The data were collected at McDonald
Observatory using the Imaging Grism Polarimeter on the Otto Struve 2.1m telescope.  Most
of the galaxies did not have significant level of linear polarization, where the bulk is $<1\%$.  A fraction of the galaxies showed a loose correlation between the polarization and position angle of the galaxy, indicating that dust scattering is the main source of optical polarization.  The unbarred spiral galaxies are consistent with the predicted relationship with inclination from scattering models of $\sim\sin^2{i}$.
  
\end{abstract}
\section{Introduction}

Measuring optical polarization is a powerful method for probing the geometry and composition of
dust in galaxies because the data contain directional and intensity
information \citep{1949Sci...109..166H,1949Sci...109..165H,2003dge..conf.....W}.  Polarization studies have been done on a
variety of objects, including
stars in the Milky Way (MW)
\citep{2000AJ....119..923H,1956ApJ...124...43V,1970MmRAS..74..139M}, accretion disks
around black holes
\citep{1984ApJ...278..499A,1984ApJ...279..485S,1992ApJ...398..454W,1990MNRAS.244..577B},
and young stellar
objects \citep{1990ApJ...364..232B}.  Measuring optical polarization of galaxies is a key way
to learn more
about their dust and magnetic field properties.
 
Studies of stars in our galaxy indicate that optical polarization arises from
directional absorption or scattering by aligned elongated dust grains in the
interstellar medium (ISM)
\citep{2003dge..conf.....W,1970MmRAS..74..139M,1976MNRAS.177..499A} commonly known
as the Davis-Greenstein effect \citep{1951ApJ...114..206D}.  These grains align
themselves with their long axis
perpendicular to the interstellar magnetic field \citep{1997ApJ...477L..25W,2003JQSRT..79..881L}.  Light
that is polarized parallel to the magnetic field, along the short axis of the dust
grain, is more likely to pass
through the ISM without being absorbed or scattered.  Thus, the observer sees linear optically polarized light that is parallel to the galactic plane.  The amount
of light that is polarized is typically between $0\%$ and $5\%$, with a few
sightlines showing polarizations as high as $12\%$, depending on the amount
and type of dust \citep{2000AJ....119..923H}.

The other source of extragalactic optical polarization is by
dust scattering \citep{1977A&A....57..141B,1980MNRAS.192...53T,1982MNRAS.200...91S}.  Light scattered off of particles will
be polarized, as in the classic example of the light scattering off Earth's
atmosphere causing the sky to be polarized.  The polarization of scattered light depends on the distribution of dust in the disk, and so the polarization angle can also be correlated with the galactic plane.  In this case the polarization would be perpendicular to the plane of the galaxy.  There have been several models of scattering in both optically thin and thick spiral galaxies and with Mie, Thomas, and Rayleigh scattering \citep{1996ApJ...465..127B,1997AJ....114.1405W,2000MNRAS.319..497S}.  They predict that the polarization should be related to the position and inclination angles, and be on the order of a few percent.  

Most studies of polarization in extragalactic sources have been done at radio
wavelengths.  Radio synchrotron emission is naturally polarized and directly traces
ordered magnetic fields in galaxies \citep{2005LNP...664...41B}.  
From extragalactic radio polarization surveys, spiral galaxies have
coherent galactic magnetic fields that lie along the galactic plane
\citep{1978PASJ...30..315T,1983IAUS..100..159B}.  Some barred spiral galaxies have
additional
magnetic field structures associated with the bar \citep{2002A&A...391...83B}.  
\citet{2009ApJ...693.1392S} conducted an integrated radio polarization survey
of $43$ galaxies and observed that the percent polarization depends on the
inclination, and the polarization angle is related to the angle of the galactic
plane.

There have been few optical polarization measurements of external galaxies, and
individual galaxies in these studies typically show optical linear polarization at
the $\lesssim 1\%$ level
\citep{1975ApJ...196..261S,1991MNRAS.252..288B,1986ARA&A..24..459S}.  The
results of these studies are mainly consistent with optical polarization arising by
the Davis-Greenstein effect and a few cases from scattering \citep{1997AJ....114.1405W,2000MNRAS.319..497S}.   

Most of the previous optical polarization studies of galaxies were spatially resolved \citep{1986ARA&A..24..459S,1990IAUS..140..245S,1995MNRAS.277.1430D}, with a few exceptions where only certain types of galaxies were targeted \citep{1990MNRAS.244..577B,1991MNRAS.252..288B}.  
\citet{1990IAUS..140..245S} and \citet{1986ARA&A..24..459S} produced optical
polarization maps of $7$ and $13$ galaxies, respectively, and found that only those with ordered dust lanes, such
as Sa type galaxies (e.g. M104) had a significant amount of
polarization, $\sim4-5\%$ along the galactic plane.  Integrated polarization should be less than what is measured from the polarization maps.  Cancellations arise from integrating the polarization
vectors over the entire galaxy.  These vectors are not well ordered,
especially at larger radii, and will partially cancel each other.
\citet{1975ApJ...196..261S} studied integrated optical polarization of $14$ galaxies.  Most of their galaxies were
ellipticals ($9$ of the $14$) and were taken without a filter. The highest
value was from NGC 7041 with a linear polarization of $0.68\pm0.12\%$, and the
average for the $14$ galaxies was $0.38\%$.  \citet{1991MNRAS.252..288B} observed starbursting and interacting galaxies in
optical/infrared polarization and found that most of them were $<1\%$
polarized.

 

Our integrated optical polarization survey of nearby galaxies is currently the largest survey
of this type, with data for 70 galaxies presented in this paper.  If dust alignment or scattering
is the main source of optical polarization, then we would expect the linear
polarization to correlate with the galaxy's position angle, morphological type, and inclination angle \citep{2000MNRAS.319..497S,1996ApJ...465..127B}. 

\section{Data Collection and Analysis} 

The data were collected at McDonald Observatory with the $2.1$m Otto Struve
telescope using the Imaging Grism Polarimeter (IGP)
\citep{1994PhDT........22T}.  Targets were selected from the Uppsala General Catalogue (UGC).  They needed
to fit within the $\sim 1 \arcmin$ diameter field of view (FOV) 
and to be brighter than $16$th magnitude \citep{1973UGC...C...0000N}.  Selected galaxies were located $>20^\circ$ off the galactic plane (with one accidental exception) and had low levels of dust extinction $E(B-V)\lesssim 0.1$, to minimize Milky Way foreground polarization \citep{1982AJ.....87.1165B,1998ApJ...500..525S}.  There was a preference for galaxies with a star $\sim0.5\arcmin$ away, close enough to be within the FOV and far enough to not intermingle with the galaxy.  The information used for target selection came from NASA's Extragalactic Database (NED) and the Uppsala General Catalogue (UGC).  A range of morphological types and
inclination angles was chosen.  Due to the size and brightness constraints, there was a selection bias towards S0 and Sa type galaxies \citep{1991trcb.book.....D}.  All of the target galaxies are thought to be
isolated with the exception of a few known to be galaxy pairs.

We conducted imaging polarimetry in the B-band over two observing runs, November 2008 and December 2009.  For each galaxy,
exposures were taken with the optical axis oriented to $0^\circ$, $45^\circ$,
$22.5^\circ$, and $67.5^\circ$ with respect to North, calibrated with the
polarized standard stars' published polarization angles.
Exposure times ranged from $100$ s to $200$ s depending on the surface brightness
of the galaxy, and each galaxy was imaged two or three times per orientation.

Standard stars from \citet{1992AJ....104.1563S} and \citet{1990AJ.....99.1243T} were observed throughout the observing runs.  During the first
observing run only three unpolarized standard stars and two polarized stars
could be used for data analysis.  (HD 251204 is a known variable according to
\citet{2006ARep...50..273A,1999AcA....49...59W,1999ApJS..125....1O})  For the
second observing run, seven unpolarized and eight polarized standard stars were
observed, two each per night with the exception of 19 December 2009 where we had observations of only one unpolarized standard.


All the images were bias corrected using IRAF, and the rest of the analysis,
including flat fielding, was done with IDL.  
The signal-to-noise ratios of the pixel with the highest counts in the reduced images were $\sim 100$.  In the IGP, there is a polarizing calcite beam
splitter, so each exposure produced two
images of an object in orthogonal polarization states, hereafter called top and
bottom.  After imaging at each of four different orientation angles an object had
eight associated images.  Images of the same galaxy  
with the same waveplate position angle were
stacked by aligning the centers of the object.  

The flux from the objects was multiplied by a 2-D Gaussian weight function $G$ that allowed us to more heavily weight the data from the bright center of
the galaxies compared to the faint edges where the signal to noise was low \citep{2000ApJ...536...79R}.
\begin{equation}
G=\frac{1}{2\pi\sigma_G^2}\exp{\Big[-\frac{1}{2\sigma_G^2}\Big((x-x_0)^2+(y-y_0)^2\Big)\Big]}.
\end{equation}
The values $x_0$ and $y_0$ correspond to the center of the galaxy.  
The standard deviation $\sigma_G$ in Equation 1 and the radius of the circle
that was
summed within were chosen to minimize the contribution of read-out noise and maximize the
signal to
noise ratio.  The radius ranged between $3\arcsec$ and $23\arcsec$ and is listed, along with $\sigma_G$, in Table \ref{gal}.  The normalized Stokes $q$ and $u$ vectors
were calculated by following the convention of \citet{1994PhDT........22T}.  The
main equations are
given below, where $I$ is the total intensity and $f_{angle,t/b}$ is the
integrated, weighted flux for either the top or bottom image with the optical axis
at the
given angle.
\begin{equation}
q=\frac{Q}{I}=\frac{1}{2}\Big(\frac{f_{0^\circ,t}-wf_{45^\circ,t}}{f_{0^\circ,t}+f_{45^\circ,t}}-\frac{f_{0^\circ,b}-wf_{45^\circ,b}}{f_{0^\circ,b}+f_{45^\circ,b}}\Big),
\end{equation}
\begin{equation}
u=\frac{U}{I}=\frac{1}{2}\Big(\frac{f_{22.5^\circ,t}-wf_{67.5^\circ,t}}{f_{22.5^\circ,t}+f_{67.5^\circ,t}}-\frac{f_{22.5^\circ,b}-wf_{67.5^\circ,b}}{f_{22.5^\circ,b}+f_{67.5^\circ,b}}\Big).
\end{equation}
The $w$ is a weighting factor to correct for changes in the flux due to differences in
atmospheric transparency, see \citet{1994PhDT........22T} for more details.  

With the normalized Stokes $q$ and $u$ vectors it is then straightforward to
calculate the linear percent of polarization and the polarization angle
(direction of the polarization E-vector):  
\begin{equation} P^2=q^2+u^2,
\end{equation} 
\begin{equation} \theta_{pol}=\frac{1}{2}\arctan{\frac{u}{q}}.
\end{equation} 


The percent polarization was corrected for a positive bias following the procedures
outlined by \citet{1997ApJ...476L..27W}.  They used a simple method for optimization of the marginal distribution \citep{1985A&A...142..100S} to find the true values $P^{true}$ and $\theta_{pol}^{true}$ by,
\begin{equation}
P^{true}=P-\sigma^2/P\;H(P-\sigma),\quad\theta^{true}_{pol}=\theta_{pol}.
\end{equation}
Here $H$ is the Heaviside function and $\sigma$ is the Poisson error based on photon statistics for the integrated polarization $P$, which includes all sources of error and is listed in Table \ref{gal}.  

The position angle of a galaxy on the sky, defined here as the projected
angle to the major axis of the brightest galactic feature from North, was determined using the second moment of the
flux $f(x,y)$ after applying the Gaussian weight function $G$ (Equation 1).  For face-on galaxies, the position angle gives the general direction of any bright internal structure, like bars and spiral arms, which from radio polarimetry traces the magnetic field \citep{2005LNP...664...41B}.  For edge-on galaxies, it traces the major axis of the the disk.  Both the polarization and position angle are calculated after applying the Gaussian weight function $G$ to allow for a direct comparison.
The position angle is given by the following equation.
  
\begin{equation}
\theta_{pa}=\frac{1}{2}\arctan{\frac{2I_{xy}}{I_{xx}-I_{yy}}}, 
\end{equation}
where, 
$$I_{xx}=\int dx\;dy\;(x-x_0)^2G\;f(x,y),$$ 
\begin{equation} I_{yy}=\int dx\;dy\;(y-y_0)^2G\;f(x,y), \end{equation} 
$$I_{xy}=\int dx\;dy\;(x-x_0)\;(y-y_0)\;G\;f(x,y).$$

The position angles calculated are all consistent with published values, with a few exceptions (i.e. \citet{1973UGC...C...0000N}).  The position angles of UGC 3709, 4227, 4556, and 5339NED01 differed from the published ones because they were calculated with the Gaussian weight function which traced the bugle instead of the disk.  For UGC 545, 4851, and 5097, the published values were used instead of the calculated position angles.  For these galaxies, with the Gaussian weight function, it was difficult to calculate decent position angles due to either stray star light from a foreground star or the inner regions were too clumpy.  For UGC 758, 1150, 1359, 2143, 3258, 3414, 3930, 11734, and 12608, position angles are published only in the infrared and deviate from our calculated angles by more than $10^\circ$. 

For the polarization percent and angle, and position angle, Poisson errors were
assumed and propagated through the calculations.  According to
\citet{1994PhDT........22T}, the instrumental polarization errors for the IGP
are $\leq 0.1\%$ which is consistent with the linear polarization values that
we measure in our polarized standard stars.  Table \ref{tbl-1} lists the
unpolarized standard star observations and Table \ref{tbl-2} lists the
polarized standard star observations with the measured and published amounts
of linear polarization for each star in the Johnson B-filter
\citep{1992AJ....104.1563S,1990AJ.....99.1243T}.

UGC 1236 and 1260, were observed during both observing runs to check for consistency.  Their measured polarization percentages and angles, as well as the calculated
position angle on the sky were in agreement for the two observing runs.  

Additionally, the inclination angle $i$ was calculated using the values of the
semi-major $a$ and semi-minor $b$ axis given in the NASA Extragalactic Database
(NED) \citep{1973UGC...C...0000N} by the following equation,

\begin{equation}
\cos^2{i}=\frac{(b/a)^2-q_0^2}{1-q_0^2}.
\end{equation}
Here, $q_0$ is the intrinsic axial ratio and for a typical spiral galaxy is $0.2$
\citep{1988ngc..book.....T}. 

There is a large amount of uncertainty in the inclination angles and the axes here are measured in unpolarized light.  They are sufficient for the purposes of this paper, since we are only concerned with the overall relationship between the polarization and inclination angle. 

\begin{figure}
\epsscale{1}\plottwo{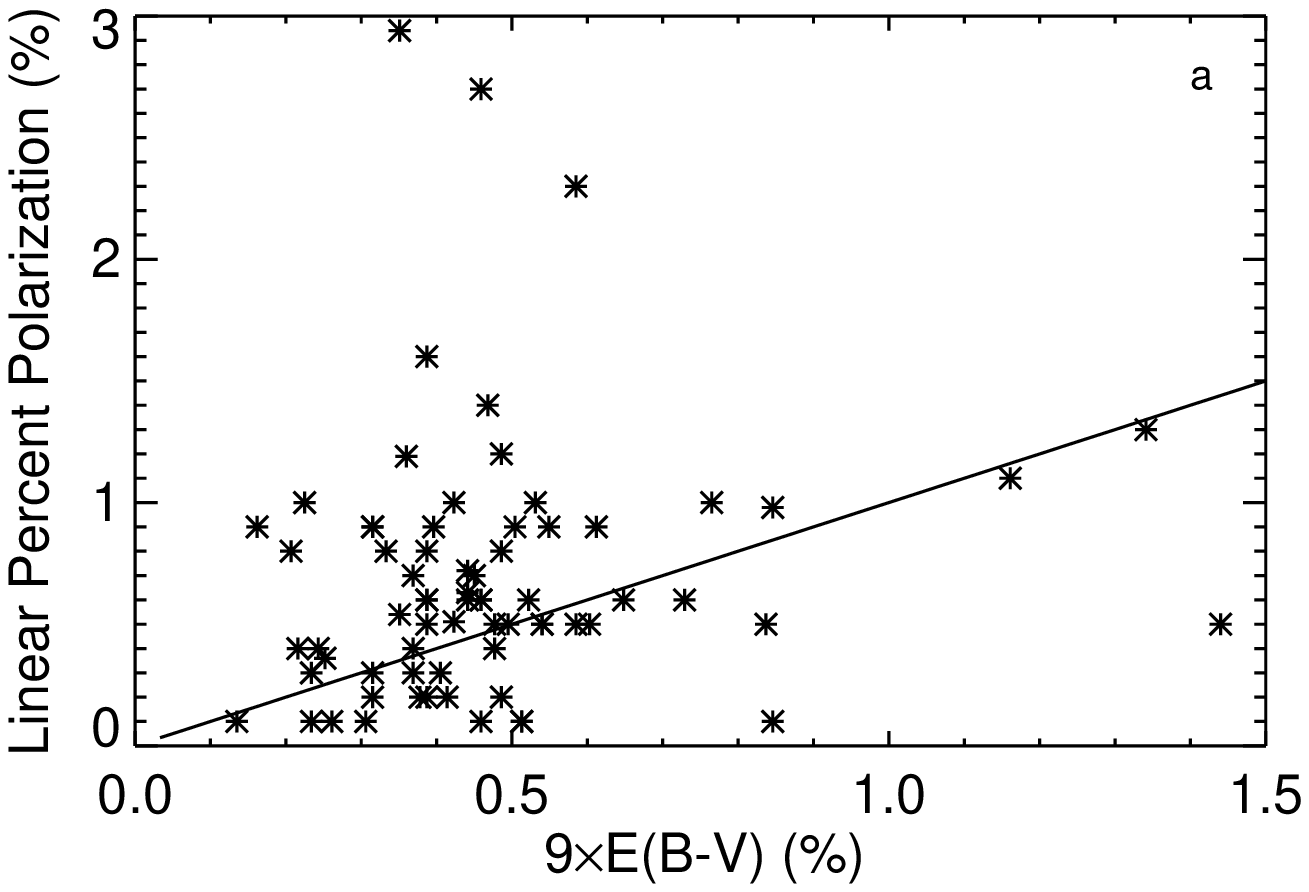}{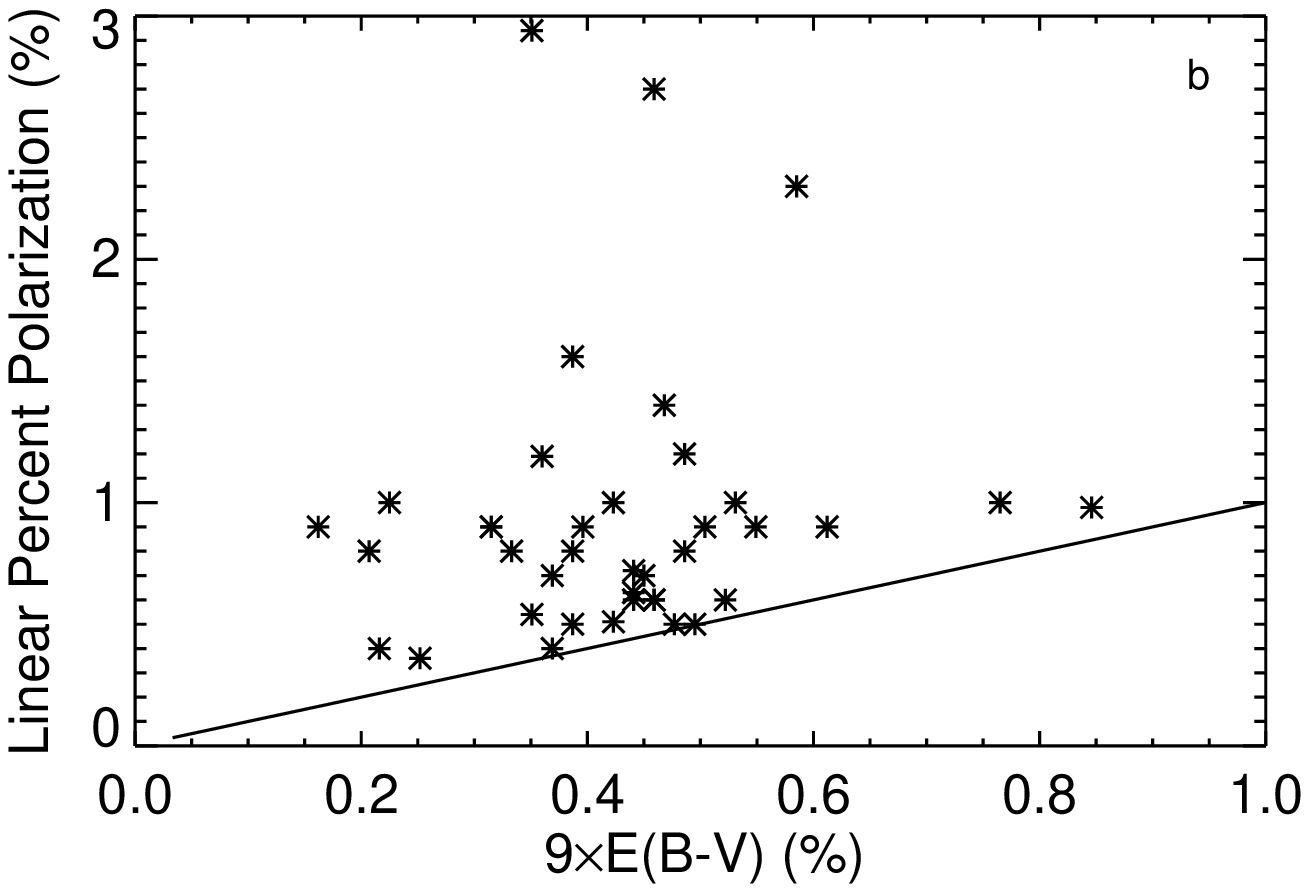}
\caption[caption]{The linear polarization (\%) versus the maximum MW foreground extinction $\sim 9\times E(B-V)$ (\%).  The left plot (a) shows all the observed galaxies and the right plot (b) shows only the accepted detections that were used in the analysis.  The galaxies used in the data analysis have levels of linear polarization that are greater than $9\times E(B-V)$ and are above the $2\sigma$ detection limit.  See $\S$ 2 for more details.}
\label{ee}
\end{figure}

An approximation of the MW foreground polarization is the color excess $E(B-V)$ \citep{1982AJ.....87.1165B}.  Note, there is about a $10\%$ error in $E(B-V)$.  This is a measure of the extinction from the MW, to the target, along the line of sight.  \citet{1975ApJ...196..261S} empirically found that the maximum optical polarization along a sightline is $\lesssim 9\times E(B-V)$.  To ensure that the polarization measured was not solely from the MW, all galaxies used in the data analysis have levels of polarization that are $\geq 9\times E(B-V)$ (Figure \ref{ee}).

UGC 3414 is a good example showing that the extinction cut is valid.  UGC 3414 and the two stars in the FOV all had similar polarization levels and angles.  Its maximum MW polarization based on the color excess $E(B-V)$ was also comparable to the measured polarization, within $1\sigma$. In this case the MW foreground contamination is the dominant source of polarization and the galaxy was removed from the analysis based on its extinction.

Table \ref{gal} presents the radius $r$ and $\sigma_G$ used in the Gaussian weight function $G$, percentage of linear polarization, polarization angle, position angle of the galaxy on the sky, as well as morphological type, MW extinction $E(B-V)$ and inclination angle from NED for the galaxies in this study \citep{1991trcb.book.....D,1982AJ.....87.1165B,1973UGC...C...0000N}.  Galaxies that were not used in the data analysis are presented below the line in Table \ref{gal}.  They either had polarizations that were less than twice the errors (similar to the cuts made by \citet{1991MNRAS.252..288B}), or
appear to be highly contaminated by MW polarization based on MW extinction $E(B-V)$.  The distributions of the polarization values before and after the cuts are shown in Figure \ref{pp}.  The majority of galaxies with $\leq0.6\%$ are not used in the analysis.  
Most of the galaxies used in the analysis are around $1\%$ linearly polarized, with a mean average
polarization of $0.94\%$. 
This is consistent with previous optical polarization studies \citep{1975ApJ...196..261S,1986ARA&A..24..459S,1990IAUS..140..245S,1991MNRAS.252..288B,1995MNRAS.277.1430D}.

\section{Discussion}



The low levels of integrated linear polarization reported in this and
previous studies may be caused by MW foreground contamination, nonuniform dust distributions inside the host galaxies, and a reduction in the integrated signal from
cancellations.  First, MW polarization is of the same order as the polarization from the bulk of the target galaxies.  Even though we have removed galaxies dominated by MW foreground polarization, the MW could still contaminate and depolarize the signal of the target galaxies.  Second, for directional absorption or scattering to occur,
the dust grains need to be be distributed somewhat uniformly about the galactic plane.  For directional absorption, a fraction of the dust grains also need to align themselves with the magnetic field \citep{2003dge..conf.....W}.  In scattering, theoretical papers assume that the dust grains are uniform and predict polarizations around 1 to 2\% \citep{1996ApJ...465..127B,1997AJ....114.1405W,2000MNRAS.319..497S}.  Lastly, the amount of
polarization could also be diminished due to cancellations from integrating over
the whole galaxy arising from non-uniformity of the galactic magnetic field.  All of these issues probably contribute to the overall low levels of polarization.

Other optical polarization studies discuss having difficulty
with the MW foreground.  In a study of optical/infrared polarization of Seyfert galaxies, \citet{1990MNRAS.244..577B} claim
that any linear polarizations less than $0.5\%$ can arise from the MW.  From
their simulation for their sample, the median degree of MW polarization was
$0.6\%$.  One of the advantages of having a large sample is the ability to avoid galaxies with a high probability of being contaminated with MW polarization.  In doing so, our initial sample size was cut in half.  Since our average polarization after the cuts is $0.9\%$, we are confident that our sample is above the noise of the MW foreground, which is $\sim 0.5\%$ based on the average $E(B-V)$.

We find a loose correlation between the position angle of the galaxy and
its integrated polarization angle, indicating that the polarization arises
from scattering off of dust grains that are distributed along the galactic plane. We examine this
correlation by looking at the distribution of the difference of these two
angles folded about $90^\circ$, shown in Figure \ref{pol}.  In the distribution we expect peaks around
$0^\circ$ for dust alignment or $90^\circ$ for scattering of the positive
difference between the polarization and position angles.  There is a peak around $90^\circ$, indicating that dust scattering is the dominant process of optical polarization.  From Monte Carlo simulations, there is less than a $10\%$ probability that this peak is random.  This finding is contradictory to what is seen in the MW optical polarization and some of the optical polarization maps of other galaxies, which are caused by dust alignment \citep{2003dge..conf.....W,1990IAUS..140..245S}.  The reason for this discrepancy is unclear. 

Galaxies with polarization levels greater than $1.5\%$ are less than
$15^\circ$ away from having their polarization being perpendicular to their position angles.  The galaxies with the highest level of polarization are consistent with scattering being the source of polarization.  The exception is UGC
545, which has the angles being parallel.  UGC 545 is a known Seyfert I galaxy according to NED \citep{1991trcb.book.....D}.  The polarization
from this galaxy is most likely due to its active galactic nucleus (AGN) and not from the galaxy.  UGC 545 has a level of polarization of $2.3\%$.  Seyfert I galaxies tend to have polarizations $\sim 2.9\%$ compared to the average of this sample of $1.0\%$, indicating that UGC 545's polarization comes from its AGN \citep{1984ApJ...278..499A}.

There is a lack of evidence from this study showing that the level of polarization depends on the morphological type.  Naively one would expect the polarization in spirals to be higher due to the coherent magnetic fields and higher amount of structure compared to irregulars and ellipticals \citep{2005LNP...664...41B,2003A&A...405..513C}.  Above $1\%$ polarization spiral galaxies are the only morphological type present, as
shown in Figure \ref{pp_pol}, however the majority of the sample is composed of spirals.  The
mean average linear polarization for spiral galaxies is $1.0\%$, 
whereas the mean average of the four irregulars is $0.81\%$ and the one elliptical galaxy is $1.0\%$.  A larger sample is needed to conclude whether there is a correlation between the percent of polarization and the morphological type of galaxies.

The polarization level of unbarred spiral galaxies generally increases with
inclination angle and barred
spirals stay roughly constant, at $\lesssim 1.0\%$, as
shown in Figure \ref{inc}.  The polarizations of unbarred spiral galaxies are predicted to vary as $\sin^2{i}$ and the galaxies in this sample loosely follow this relationship \citep{1996ApJ...465..127B,2000MNRAS.319..497S}.  A similar trend was found in an integrated radio
polarization survey \citep{2009ApJ...693.1392S}.  Galactic magnetic fields are
toroidal \citep{2002ChJAA...2..293H}, so less inclined galaxies should have more of their
polarization cancel out from the integration.  However, bars in spiral galaxies have
been shown to have additional magnetic fields associated with them
\citep{2005LNP...664...41B}, so a face-on barred spiral could have a higher polarization level due to the contribution from the bar.

\section{Summary}
 
We have presented the results of an integrated optical polarization survey of
70 nearby galaxies in order to study the relationship between linear polarization
and galactic properties.

\begin{figure}
\epsscale{1}\plotone{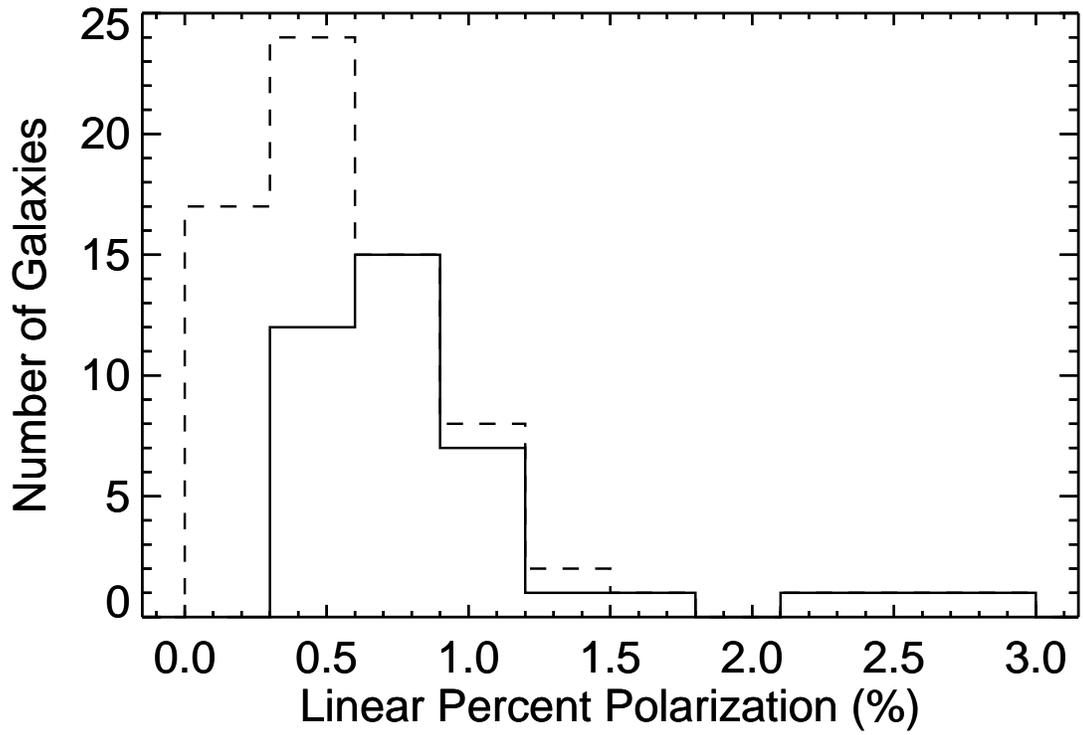}
\caption[caption]{A histogram of the distributions of the linear polarization (\%),
with bin size of $0.3\%$.  All the observed galaxies (dashed line) peak around $0.5\%$.  After the extinction and error cuts ($\S$ 2), the galaxies that are used in the analysis (solid line) have an average polarization of $0.9\%$.}
\label{pp}
\end{figure}

$\quad\bullet\quad$The average level of linear optical polarization of galaxies in this survey that were used in the analysis is $0.9\%$ (Figure \ref{pp}).  Seven of the galaxies that were used for the analysis
had polarization greater than $1\%$.  This is consistent with previous optical
polarization studies of galaxies.

\begin{figure}
\epsscale{1}\plotone{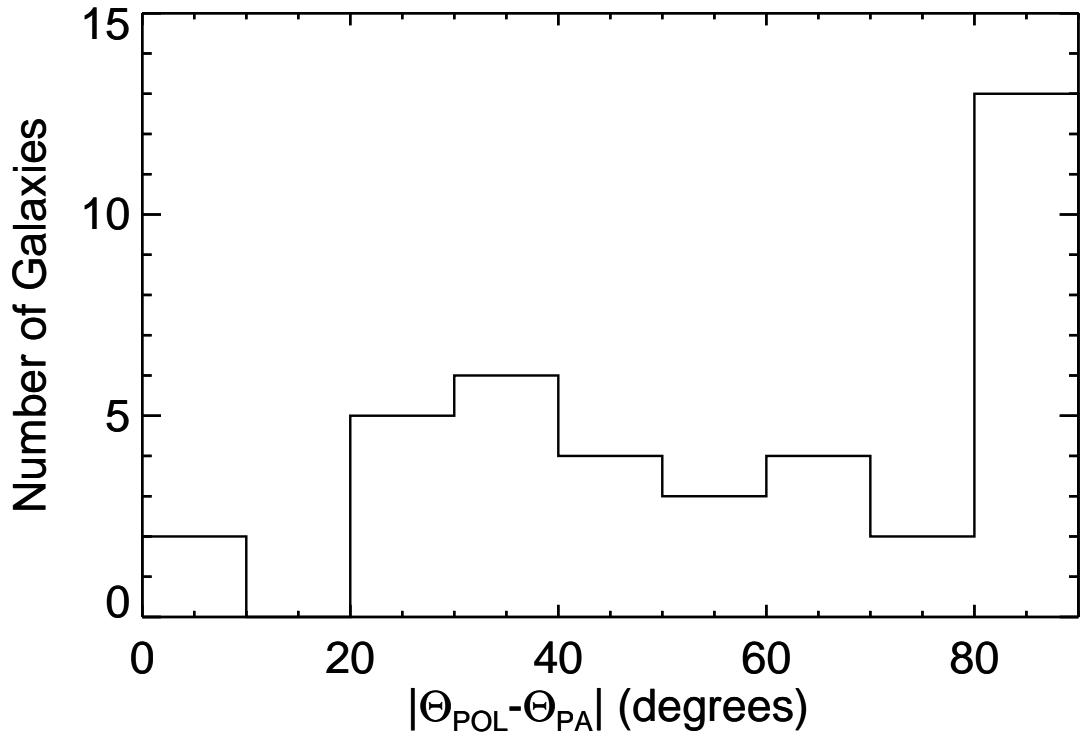}
\caption[caption]{A histogram of the distribution of the positive difference between
the polarization and position angles, with bin size of $10^\circ$.  The difference was then folded about $90^\circ$.  The peak between $80$ and $90^\circ$ means that the polarization angle is perpendicular to the position angle of the galaxy.  This suggests that scattering is the dominant source of optical polarization in these galaxies.}
\label{pol}
\end{figure}

$\quad\bullet\quad$There is a loose correlation between the galactic polarization
angle and the position angle (Figure \ref{pol}).  There is a peak in the
distribution of the positive difference between the polarization and position angle
around 90 degrees indicating that the polarization is perpendicular to the
galactic plane.  This suggests that the galactic optical polarization arises from scattering.  From Monte Carlo simulations, there is less than a $10\%$ chance that this peak is random.  

\begin{figure}
\epsscale{1}\plotone{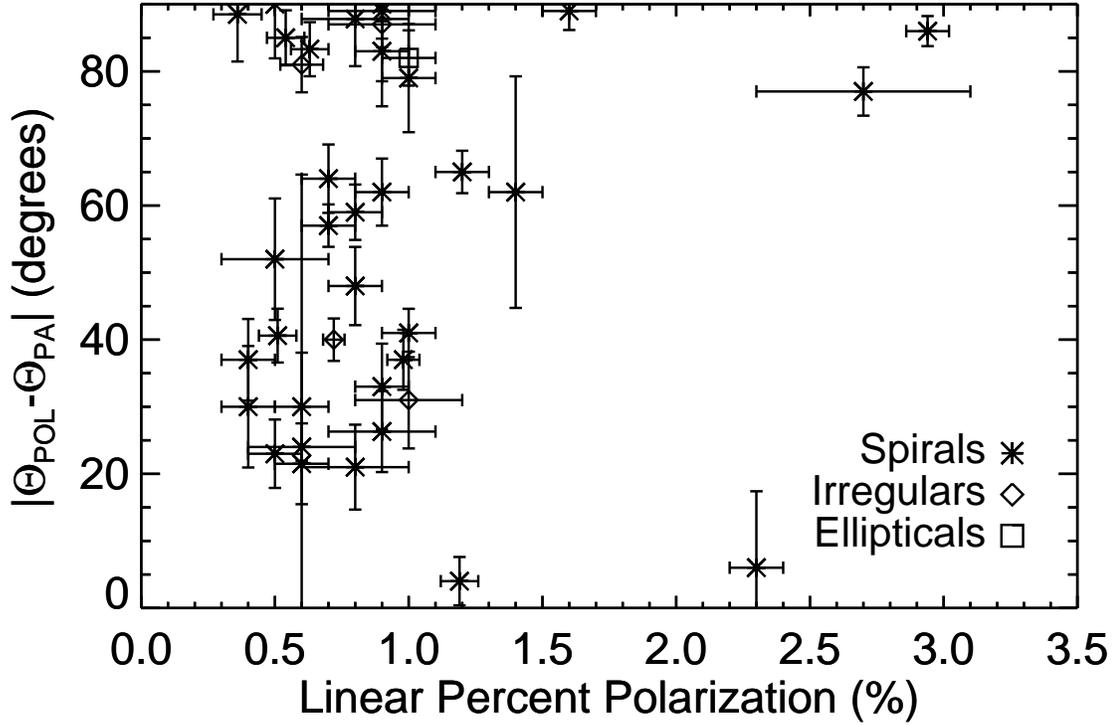}
\caption[caption]{A plot of the linear polarization versus the positive difference
between the polarization and position angle, with $1\sigma$ error bars.  Spirals are
the stars, irregulars are the diamonds, and the one elliptical is a square.  There is a high amount of scatter in the
angle below $1\%$ polarization.  The highly polarized galaxies, above $1.5\%$, all have the difference between their polarization and position angles between $77$ and $90^\circ$, indicating that their polarizations are caused by scattering.  The exception is UGC 545, a known Seyfert 1 galaxy, which is parallel (around $0^\circ$).}
\label{pp_pol}
\end{figure}

$\quad\bullet\quad$There is no evidence that the level of polarization depends on morphological type (Figure \ref{pp_pol}).

\begin{figure}
\epsscale{1}\plotone{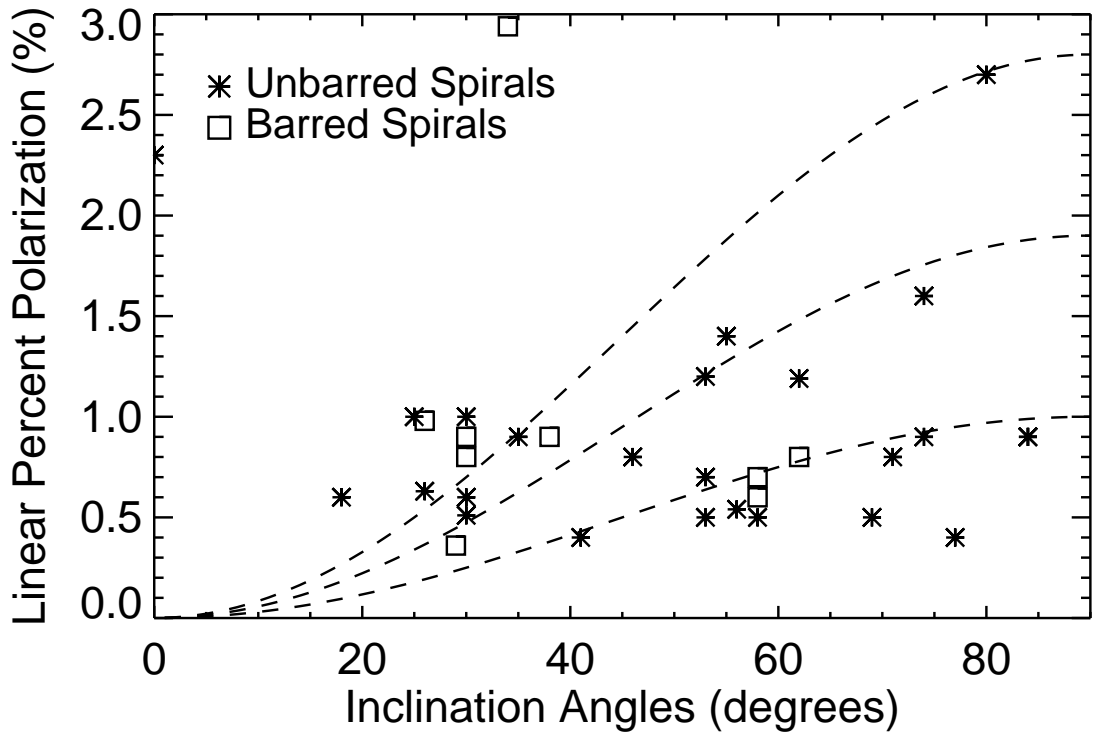}
\caption[caption]{A plot of the linear polarization as a function of inclination
angle.  The squares are the barred spirals and the stars are the unbarred
spirals.  The barred spirals tend to have a flat distribution compared
with the unbarred spirals that peak at high inclinations.  Overplotted are the theoretically predictions based on Thomson scattering for $P\sim\sin^2i$ \citep{2000MNRAS.319..497S}.  The unbarred galaxies roughly follow these predictions.  A similar result was seen with radio polarization by
\citet{2009ApJ...693.1392S}.}
\label{inc}
\end{figure}

$\quad\bullet\quad$For unbarred spiral galaxies the polarization tends to increase
with inclination angle loosely following the predicted $\sin^2{i}$ relationship from scattering \citep{1996ApJ...465..127B,2000MNRAS.319..497S}.  Barred spirals have a fairly flat distribution (Figure \ref{inc}).  This is consistent with the \citet{2009ApJ...693.1392S} study
using radio polarization.

We thank McDonald Observatory for the access to the Otto Struve 2.1m telescope.  We would also like to thank the anonymous referee for the detailed suggestions.  This research has made use of the NASA/IPAC extragalactic database
(NED), which is operated by the Jet Propulsion Laboratory, Caltech, under contract with the National Aeronautics and Space Administration.  This work was supported by Texas A\&M University funds made available to Lifan Wang.


\clearpage
\begin{deluxetable}{cccc}
\tablecaption{Unpolarized Standard Stars\label{tbl-1}}
\tablehead{
\colhead{Name} &
\colhead{Night Observed} &
\colhead{Measured Polarization(\%)} &  
\colhead{Published Polarization (\%)}
}
\startdata
HD 94851  &11-28-08& 0.080 $\pm$ 0.08 & 0.06 $\pm$ 0.02  \\
HD 94851 & 11-29-08 & 0.14 $\pm$ 0.08 & 0.06 $\pm$ 0.02 \\
BD +28$^\circ$4211 & 11-30-08 & 0.18 $\pm$ 0.05 & 0.063 $\pm$ 0.023 \\
HD12021\tablenotemark{a}  & 12-17-09 &0.31 $\pm$ 0.03 & 0.112 $\pm$ 0.025 \\
GD319  & 12-17-09 & 0.19 $\pm$ 0.1 & 0.045 $\pm$ 0.047 \\
HD12021\tablenotemark{a}  & 12-18-09 & 0.28 $\pm$ 0.05 & 0.112 $\pm$ 0.025  \\
HD14069  & 12-18-09 & 0.27 $\pm$ 0.1 & 0.111 $\pm$ 0.036   \\
HD94851  & 12-19-09 & 0.0039 $\pm$ 0.05 & 0.06 $\pm$ 0.02  \\
HD14069  & 12-20-09 & 0.31 $\pm$ 0.05 & 0.111 $\pm$ 0.036 \\
GD319  & 12-20-09 & 0.13 $\pm$ 0.15 & 0.045 $\pm$ 0.047 
\enddata
\tablenotetext{a}{HD 12021 is consistently higher than the published value and is
likely no longer an acceptable unpolarized standard star.}
\tablecomments{All published values are from \citet{1992AJ....104.1563S}, except for
HD 94851 which is from \citet{1990AJ.....99.1243T}.  }
\end{deluxetable}

\clearpage
\begin{deluxetable}{cccc}
\tablecaption{Polarized Standard Stars\label{tbl-2}}
\tablehead{
\colhead{Name} & 
\colhead{Night Observed} &
\colhead{Measured Polarization(\%)} &  
\colhead{Published Polarization (\%)}
}
\startdata
HD 245310  &11-29-08 & 4.5 $\pm$ 0.05 & 4.550 $\pm$ 0.064 \\
HD 251204\tablenotemark{a} & 11-29-08 & 4.8 $\pm$ 0.08 & - \\
BD +38$^\circ$4058/Hilt 960  &11-30-08& 5.6 $\pm$ 0.1 & 5.648 $\pm$ 0.022\\
BD +59$^\circ$ 389  &12-17-09& 6.3 $\pm$ 0.07  & 6.345 $\pm$ 0.035 \\
HD245310  &12-17-09& 4.6 $\pm$ 0.09  & 4.550 $\pm$ 0.064 \\
BD +64$^\circ$106  &12-18-09& 5.7 $\pm$ 0.07  & 5.506 $\pm$ 0.090 \\
HD245310  & 12-18-09 & 4.6 $\pm$ 0.1  & 4.550 $\pm$ 0.064  \\
BD+64$^\circ$106  &12-19-09 & 5.6 $\pm$ 0.07 & 5.506 $\pm$ 0.090 \\
HD245310  &12-19-09 & 4.5 $\pm$ 0.09  & 4.550 $\pm$ 0.064 \\
BD +64$^\circ$106  &12-20-09 &  5.6 $\pm$ 0.07 & 5.506 $\pm$ 0.090 \\
HD245310  &12-20-09 & 4.5 $\pm$ 0.08  & 4.550 $\pm$ 0.064 \\
\enddata
\tablenotetext{a}{HD 251204 is a variable according to
\citet{2006ARep...50..273A,1999AcA....49...59W,1999ApJS..125....1O} and is not an
acceptable polarized standard star.}
\tablecomments{All published values, except HD 251204, are from
\citet{1992AJ....104.1563S}.}
\end{deluxetable}

\clearpage
\begin{deluxetable}{ccccccccc}
\tabletypesize{\scriptsize}
\tablecaption{Galaxies\label{gal}}
\tablehead{
\colhead{UGC} &
\colhead{$r$ ($\arcsec$)} &
\colhead{$\sigma_G$ ($\arcsec$)} &
\colhead{Linear} &  
\colhead{Polarization} &
\colhead{Position} &
\colhead{Morphological} &
\colhead{$E(B-V)$} &
\colhead{Inclination} \\
\colhead{} & 
\colhead{} & 
\colhead{} & 
\colhead{Polarization(\%)} &  
\colhead{Angle ($^\circ$)} &
\colhead{Angle ($^\circ$)} &
\colhead{Type} &
\colhead{} &
\colhead{Angle ($^\circ$)}
}
\startdata
6 & 8.7 & 1.9 &1.0 $\pm$ 0.2 & 122 $\pm$ 4 & 91 $\pm$ 6 & pec & 0.047 & $\ldots$\\
50 & 6.8 & 3.1 &0.9 $\pm$ 0.2 & 100 $\pm$ 4 & 11 $\pm$ 1 & Sab & 0.035 & 74\\
108 & 11.2 & 5.0 &0.7 $\pm$ 0.1 & 122 $\pm$ 5 & 58 $\pm$ 1 & SBb & 0.041 & 58\\
240 & 14.3 & 5.6 &0.8 $\pm$ 0.1 & 13 $\pm$ 4 & 72 $\pm$ 1 & SAB(rs)b & 0.023 & 62\\
285 & 7.4 & 4.3 &0.4 $\pm$ 0.1 & 138 $\pm$ 9 & 108 $\pm$ 1 & Sa & 0.041 & 77\\
425 & 11.8 & 5.0 &2.9 $\pm$ 0.1 & 135 $\pm$ 1 & 41 $\pm$ 2 & SB & 0.039 & 34\\
529 & 9.3 & 3.1 &0.9 $\pm$ 0.2 & 121 $\pm$ 6 & 95 $\pm$ 1 & S0/a & 0.061 & 84\\
540 & 19.8 & 5.6 &0.5 $\pm$ 0.1 & 107 $\pm$ 5 & 130 $\pm$ 1 & Sb & 0.053 & 53\\
545 & 5.6 & 1.2 &2.3 $\pm$ 0.1 & 144 $\pm$ 3 & 150 $\pm$ 10\tablenotemark{b} & Sa & 0.065 & 0\\
555 & 8.7 & 2.5 &0.9 $\pm$0.1 & 118 $\pm$ 5 & 28 $\pm$ 4 & S0/a & 0.056 & 84\\
697 & 16.7 & 5.0 &0.6 $\pm$ 0.1 & 100 $\pm$ 6 & 79 $\pm$ 1 & SB & 0.049 & 58\\
796 & 12.4 & 2.5 &0.6 $\pm$ 0.1 & 11 $\pm$ 4 & 41 $\pm$ 7 & S & 0.058 & 30\\
1086 & 10.5 & 3.7 &0.5 $\pm$ 0.1 & 138 $\pm$ 8 & 48 $\pm$ 1 & S & 0.043 & 69\\
1220 & 13.6 & 5.6 &1.2 $\pm$ 0.1 & 119 $\pm$ 3 & 4 $\pm$ 1 & Spec & 0.054 & 53\\
1236\tablenotemark{a} & 9.9 & 2.5 & 0.8 $\pm$ 0.2 & 102 $\pm$ 6 & 81 $\pm$ 2 & S & 0.043 & 46\\
1251 & 9.9 & 2.5 &0.5 $\pm$ 0.2 & 92 $\pm$ 9 & 40 $\pm$ 1 & S/Irr & 0.055 & 58\\
1359 & 13.6 & 3.7 &0.8 $\pm$ 0.1 & 109 $\pm$ 5 & 61 $\pm$ 3 & SB/Sb & 0.054 & 30\\
1467 & 12.4 & 5.6 & 0.7 $\pm$ 0.1 & 142 $\pm$ 3 & 85 $\pm$ 1 & S & 0.050 & 53\\
2143 & 11.2 & 5.0 &0.6 $\pm$ 0.1 & 126 $\pm$ 4 & 45 $\pm$ 1 & Irr & 0.051 & $\ldots$\\
3258 & 3.1 & 1.9 &0.9 $\pm$ 0.1 & 131 $\pm$ 5 & 98 $\pm$ 4 & SB(r)ab pec & 0.068 & 30\\
3885 & 17.4 & 7.4 &0.6 $\pm$ 0.1 & 167 $\pm$ 4 & 84 $\pm$ 1 & S & 0.049 & 26\\
3930 & 13.0 & 5.0 &0.7 $\pm$ 0.1 & 159 $\pm$ 3 & 119 $\pm$ 1 & Im & 0.049 & $\ldots$\\
4028 & 18.6 & 8.1 &0.4 $\pm$ 0.1 & 108 $\pm$ 7 & 20 $\pm$ 1 & SAB(s)c & 0.028 & 29\\
4079 & 11.8 & 2.5 &1.2 $\pm$ 0.1 & 13 $\pm$ 3 & 9 $\pm$ 2 & Spec & 0.040& 62\\
4229 & 5.6 & 1.2 &0.6 $\pm$ 0.2 & 170 $\pm$ 7 & 146 $\pm$ 40 & S0 & 0.051 & 18\\
4249 & 8.7 & 4.3 &2.7 $\pm$ 0.4 & 158 $\pm$ 3 & 55 $\pm$ 2 & S &0.051 & 80\\
4438 & 15.5 & 6.8 &0.5 $\pm$ 0.1 & 152 $\pm$ 4 & 13 $\pm$ 1 & S(r) pec & 0.047 & 30\\
4709 & 11.2 & 5.0 &0.4 $\pm$ 0.1 & 171 $\pm$ 6 & 28 $\pm$ 1 & S & 0.024 & 41\\
4730 & 7.4 & 1.9 &1.6 $\pm$ 0.1 & 7 $\pm$ 2 & 98 $\pm$ 2 & S & 0.043 & 74\\
4737 & 6.8 & 1.9 &1.4 $\pm$ 0.1 & 50 $\pm$ 3 & 168 $\pm$ 20 & S0 & 0.052 & 55\\
4813 & 7.4 & 2.5 &0.9 $\pm$ 0.1 & 78 $\pm$ 4 & 161 $\pm$ 2 & S & 0.018 & 35\\
5097 & 11.2 & 3.7 &0.5 $\pm$ 0.1 & 150 $\pm$ 4 & 55 $\pm$ 1\tablenotemark{b} & Sb & 0.039 & 56\\
5132 & 11.2 & 5.0 &0.8 $\pm$ 0.2 & 121 $\pm$ 7 & 29 $\pm$ 1 & S & 0.037 & 71\\
5200 & 8.1 & 1.9 &1.0 $\pm$ 0.1 & 43 $\pm$ 4 & 144 $\pm $ 7 & S0 & 0.025 & 30\\
5339 NED02 & 5.6 & 1.9 &0.9 $\pm$ 0.2 & 11 $\pm$ 7 & 104 $\pm$ 10 & Gpair & 0.035 & $\ldots$\\ 
11734 & 21.7 & 8.7 &1.0 $\pm$ 0.1 & 83 $\pm$ 4 & 46 $\pm$ 2 & SBb & 0.094 & 26\\
12329 & 9.9 & 3.1 &1.0 $\pm$ 0.1 & 123 $\pm$ 4 & 25 $\pm$ 1 & E3pec & 0.085 & $\ldots$\\
12608 & 16.7 & 5.6 &1.0 $\pm$ 0.1 & 61 $\pm$ 3 & 102 $\pm$ 2 & SA(r)bc & 0.059 & 25\\
NGC 7629 & 6.2 & 1.9 &0.9 $\pm$ 0.1 & 120 $\pm$ 4 & 2 $\pm$ 3 & SBO & 0.044 & 38\\
\tableline
644 & 9.3 & 4.3 &0.6 $\pm$ 0.3 & 144 $\pm$ 10 & 69 $\pm$ 1 & Gpair & 0.043 & $\ldots$\\
644 NOTES01 & 6.2 & 3.1 &0.6 $\pm$ 0.3 & 105 $\pm$ 10 & 18 $\pm$ 1 & Gpair & 0.043 & $\ldots$\\
645 & 7.4 & 2.5 & 0.3 $\pm$ 0.1& 115 $\pm$ 10 & 41 $\pm$ 20 & SB pec & 0.045 & 62\\
758 & 18.0 & 6.8 &0.4 $\pm$ 0.1 & 115 $\pm$ 4 & 83 $\pm$ 1 & S & 0.053 & 0\\
912 & 9.9 & 1.9 &0.5 $\pm$ 0.1 & 110 $\pm$ 6 & 115 $\pm$ 1 & $\ldots$ & 0.060 & $\ldots$\\
1135 & 11.8 & 3.1 & 0.2 $\pm$ 0.2 & 93 $\pm$ 20 & 34 $\pm$ 6 & S & 0.042 & 53\\
1150 & 19.2 & 6.2 & 0.5 $\pm$ 0.1 & 106 $\pm$ 4 & 111 $\pm$ 4 & pec & 0.067 & $\ldots$ \\
1260\tablenotemark{a} & 14.9 & 4.3 & 0.5 $\pm$ 0.1 & 108 $\pm$ 6 & 45 $\pm$ 1 & (R')SB(s)a & 0.093 & 42\\
1757 & 9.9 & 2.5 & 0.1 $\pm$ 0.1 & 161 $\pm$ 30 & 78 $\pm$ 1 & S & 0.057 & 81\\
2103 & 13.0 & 3.7 &1.3 $\pm$ 0.1 & 150 $\pm$ 2 & 9 $\pm$ 1 & S LIRG & 0.149 & 40\\
2608 & 15.5 & 3.7 &0.5 $\pm$ 0.1 & 144 $\pm$ 4 & 81 $\pm$ 1 & (R')SB(s)b & 0.160 & 28\\
3414 & 11.8 & 3.7 &1.1 $\pm$ 0.1 & 170 $\pm$ 3 & 176 $\pm$ 1 & S & 0.129 & 0\\
3709 & 22.9 & 10.5 &0.6 $\pm$ 0.1 & 172 $\pm$ 3 & 57 $\pm$ 1 & S pec & 0.072 & 22\\
3816 & 9.3 & 2.5 &0.5 $\pm$ 0.1 & 178 $\pm$ 4 & 109 $\pm$ 4 & S0 & 0.060 & 44\\
3969 & 11.2 & 4.3 & 0.1 $\pm$ 0.2 & 144 $\pm$ 50 & 135 $\pm$ 1 & Scd & 0.051 & 90\\
4041 & 15.5 & 3.7 & 0.3 $\pm$ 0.2 & 163 $\pm$ 20 & 127 $\pm$ 1 & E & 0.026 & $\ldots$\\
4111 & 18.0 & 6.2 & 0.2 $\pm$ 0.1 & 84 $\pm$ 8 & 173 $\pm$ 1 & Sbcpec & 0.046 & 44\\
4227 & 16.1 & 5.0 & 0.2 $\pm$ 0.1 & 149 $\pm$ 10 & 120 $\pm$ 3 & SAB(rs)b & 0.054 & 24\\
4287 & 16.7 & 3.1 &  0.3 $\pm$ 0.1 & 87 $\pm$ 6 & 174 $\pm$ 2 & (R)SB(r)ab & 0.041 & 28\\
4327 & 21.1 & 4.3 & 0.1 $\pm$ 0.1 & 130 $\pm$ 10 & 62 $\pm$ 1 & (R')SB(s)a & 0.026 & 44\\
4556 & 11.8 & 5.0 & 0.1 $\pm$ 0.2 & 177 $\pm$ 90 & 46 $\pm$ 1 & Irr & 0.029 & $\ldots$\\
4564 & 8.1 & 3.1 & 0.3 $\pm$ 0.2 & 0 $\pm$ 10 & 33 $\pm$ 1 & Sc(f) & 0.035 & 77\\
4580 & 13.0 & 5.0 & 0.1 $\pm$ 0.1 & 172 $\pm$ 30 & 11 $\pm$ 1 & Sbc & 0.034 & 62\\
4642 & 13.6 & 5.0 &0.4 $\pm$ 0.2 & 174 $\pm$ 9 & 35 $\pm$ 1 & S & 0.027 & 32\\
4687 & 9.9 & 1.9 & 0.1 $\pm$ 0.1 & 113 $\pm$ 20 & 110 $\pm$ 10 & pec & 0.094 & $\ldots$\\
4741 & 13.0 & 5.0 &0.5 $\pm$ 0.1 & 130 $\pm$ 3 & 67 $\pm$ 1 & S & 0.065 & 53\\
4823 & 11.2 & 4.3 & 0.2 $\pm$ 0.2  & 92 $\pm$ 20 & 173 $\pm$ 1 & Sab & 0.043 & 38\\
4851 & 7.4 & 2.5 & 0.1 $\pm$ 0.1& 96 $\pm$ 20 & 150 $\pm$ 3\tablenotemark{b} & S0- & 0.015 & 56\\
5022 & 9.3 & 2.5 & 0.1 $\pm$ 0.1 & 150 $\pm$ 40 & 177 $\pm$ 1 & Sa & 0.057 & 69\\
5339 NED01 & 7.4 & 2.5 & 0.2 $\pm$ 0.2 & 98 $\pm$ 20 & 26 $\pm$ 1 & Gpair & 0.035& $\ldots$ \\
12906 & 12.4 & 3.1 &0.6 $\pm$ 0.1 & 78 $\pm$ 4 & 73 $\pm$ 1 & S0/a & 0.081 & 55\\
\enddata
\tablenotetext{a}{This galaxy was observed during both observing runs to check for consistency.  The average values over both observing runs are quoted here.}
\tablenotetext{b}{The position angle quoted in NED was used instead of the calculated value.}
\tablecomments{$\;$Listed here for each galaxy are the radius $r$ and $\sigma_G$ used in the Gaussian weight function $G$ (Equation 1), linear percent polarization, polarization angle, position angle, morphological type, MW extinction $E(B-V)$, and the inclination angle.  The morphological types and Milky Way extinction $E(B-V)$ are from NASA's Extragalactic Database (NED) \citep{1991trcb.book.....D,1982AJ.....87.1165B}.  The inclination angles were derived from NED (Equation 9) \citep{1973UGC...C...0000N}.  All data below the line were not used in the data analysis (for details please see text).}
\end{deluxetable}

\end{document}